\documentclass[amsmath,amssymb,aps,prx,reprint,superscriptaddress]{revtex4-2}

\usepackage{amsmath}
\usepackage{xcolor}
\usepackage{amsfonts,amssymb}
\usepackage{graphicx}
\usepackage{hyperref}
\usepackage{academicons}
\usepackage{multirow}
\usepackage{mhchem}
\usepackage{booktabs} 
\usepackage[english]{babel}
\usepackage{tabularx}

\begin{document}

\title{Machine-Learning-Guided Insights into Solid-Electrolyte Interphase Conductivity: Are Amorphous Lithium Fluorophosphates the Key?}

\author{\text{Peichen Zhong}}
\email[]{zhongpc@berkeley.edu}
\affiliation{Bakar Institute of Digital Materials for the Planet, UC Berkeley, California 94720, United States}
\affiliation{Materials Sciences Division, Lawrence Berkeley National Laboratory, California 94720, United States}
\affiliation{Department of Materials Science and Engineering, National University of Singapore, Singapore 117575, Singapore}

\author{\text{Kristin A. Persson}}
\email[]{kristinpersson@berkeley.edu}
\affiliation{Bakar Institute of Digital Materials for the Planet, UC Berkeley, California 94720, United States}
\affiliation{Materials Sciences Division, Lawrence Berkeley National Laboratory, California 94720, United States}

\date{\today}

\begin{abstract}
Despite decades of study, the identity of the dominant \ce{Li+}-conducting phase within the inorganic SEI of Li-ion batteries remains unresolved. While the mosaic model describes LiF/\ce{Li2O}/\ce{Li2CO3} nanocrystallites within a disordered matrix, these crystalline phases inherently offer limited ionic conductivity. Growing evidence suggests that interfaces, grain boundaries, and amorphous phases may instead host the primary fast-ion pathways. Using diffusion-based generative structure prediction and machine-learning interatomic potentials (MLIPs), we investigate lithium difluorophosphate (\ce{LiPO2F2}), a key mixed-anion decomposition product of phosphorus- and fluorine-containing electrolytes. We identify a stable crystalline polymorph and demonstrate that the amorphous counterpart is conductive, with projected room-temperature $\sigma \approx 0.18$ mS cm$^{-1}$ and $E_\mathrm{a} \approx 0.40$ eV. This enhancement stems from structural disorder flattening the Li site-energy landscape and a low formation energy for Li-interstitial defects, which supplies additional mobile carriers. We propose amorphous mixed-anion Li--P--O--F phases as a promising conducting medium in the SEI, offering a specific target for engineering improved battery interfaces.
\end{abstract}


\maketitle

Functional surface passivation, such as the solid electrolyte interphase (SEI) in Li-ion batteries and protective oxides on stainless steels, forms spontaneously yet critically governs performance through its complex chemistry and defect kinetics. For Li-ion anodes, the inner SEI is often described as a mosaic of crystalline inorganic phases (\ce{Li2CO3}, \ce{Li2O}, \ce{LiF}) embedded in a disordered, organic/inorganic matrix \cite{peled_reviewsei_2017, Zhang2017, Oyakhire2022}. However, these well-characterized crystalline materials are fundamentally poor ion conductors, exhibiting ionic conductivities that are orders of magnitude too low to sustain the Li-ion flux required for high-rate battery operation \cite{he_intrinsic_2020, guo_li_2020, liu_probing_2025}. Specifically, their high formation energies for charge-carrying defects, such as Li interstitials \cite{yildirim_first-principles_2015, ma_origin_2022, christensen_mathematical_2004}, preclude significant extrinsic conductivity \cite{shi_defect_2013}. 
Together, these facts indicate that the dominant Li-ion transport pathways likely reside in other, as yet uncharacterized, components of the SEI.

Interfaces and grain boundaries have been proposed to provide markedly faster ionic diffusion \cite{ma_origin_2022}. Several modeling and simulation studies have shown the possibility of enhanced conduction at atomically sharp crystal grain interfaces \cite{pan_design_2016, Hu2023_jacs}, such as LiF/\ce{Li2CO3}. However, the evidence is inconclusive. 
\citet{peled_advanced_1997} proposed a circuit model for the mosaic SEI formed by a LiI-P(EO)$_n$-\ce{Al2O3} composite that included grain boundaries and concluded from experimental impedance data that the ionic resistance of these boundaries was significantly larger than that of their bulk counterparts.
Experimentally, the evidence for the importance of grain boundaries in carrier transport is indirect. 
For example, attempts to engineer the SEI to enable rapid conduction pathways using electrolytes made up of multiple salts have led to more boundaries between polymer and inorganic compounds and improved conduction \cite{wang_high_2023}. 

Interestingly, several experimental works have demonstrated that using \ce{LiPO2F2} as an electrolyte additive consistently leads to substantial enhancements in battery performance and stability \cite{chen_outstanding_2018, Li2022_LiPO2F2, tan_additive_2022}. 
Furthermore, state-of-the-art cryogenic transmission electron microscopy (cryo-TEM) revealed that while regular concentration electrolytes form a mosaic SEI with nanocrystalline domains of LiF and \ce{Li2O} embedded within a disordered matrix, high-concentration electrolytes form a homogeneous amorphous inorganic matrix \cite{Han2021_cryoTEM, chen_origin_2023}. 
Importantly, recent cryogenic X-ray photoelectron spectroscopy (cryo-XPS) suggests that lithium fluorophosphates (LiPO$_x$F$_y$), decomposition products of the \ce{LiPF6} electrolytes, constitute the primary inorganic component in the native SEI, a constituent that readily decomposes to LiF and releases \ce{POF3} under standard XPS pre-sample treatments \cite{nguyen2025_insitu}. Concurrently, cryo-XPS has characterized SO$_x$F$_y$ components as decomposition products of the LiFSI salt \cite{shuchi_cryogenic_2025}.
In addition, the concept of high single-ion conductivity in amorphous mixed-anion (e.g., Li oxychlorides \cite{zhang_family_2023}) or polyanion (e.g., Li-P-S \cite{hayashi_preparation_2001}) systems is well-established in solid electrolytes. 
Hence, several independent sources suggest that amorphous, mixed-anion lithium fluorophosphates, whether a minority grain boundary component or a majority phase, warrant a deeper investigation as candidates for the main ionic transport medium of the SEI.

\begin{figure*}[t]
    \centering
    \includegraphics[width=0.99\linewidth]{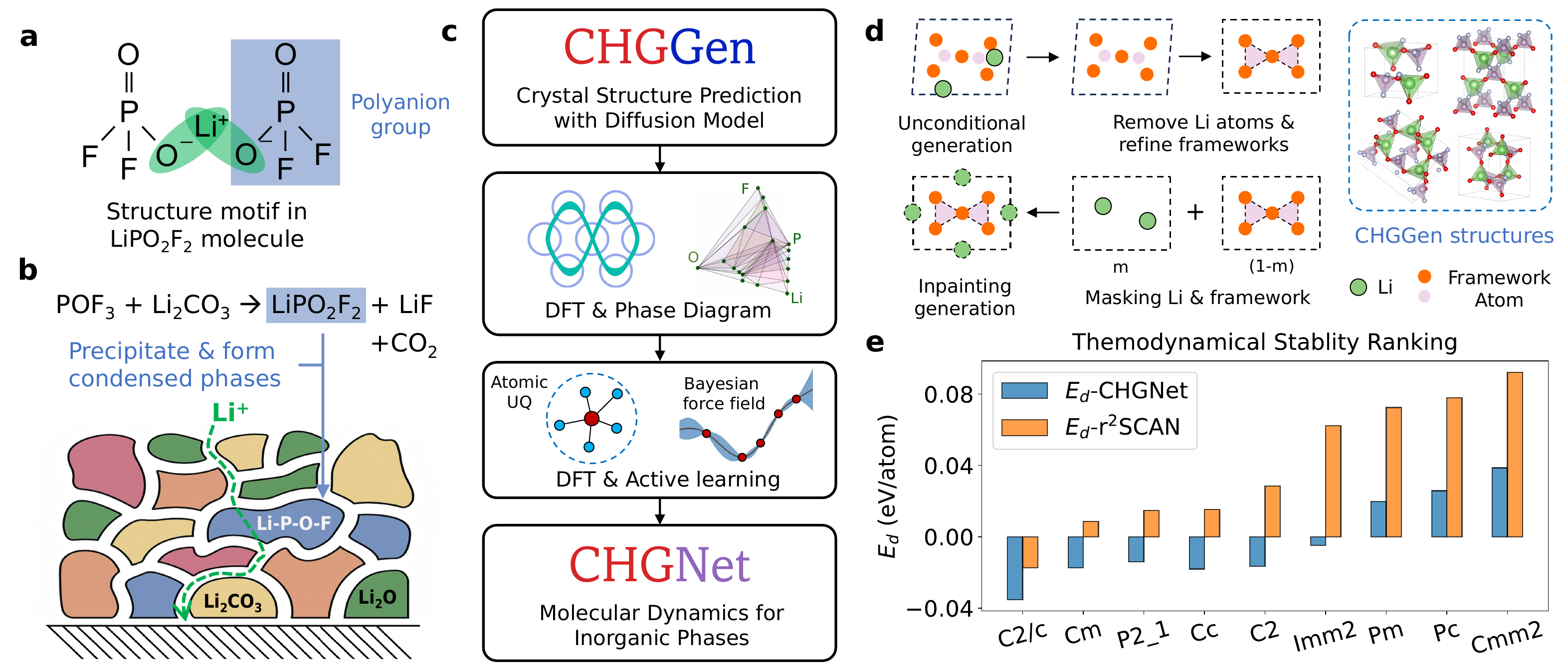}
    \caption{
        \textbf{Computational framework for crystal structure prediction and analysis of \ce{LiPO2F2}.} (a) The structural motif from the \ce{LiPO2F2} molecule, where Li atoms interconnect to form an inorganic condensed phase. (b) Schematic illustration of reaction product precipitation, inorganic SEI component distribution, and the proposed Li transport mechanism. 
        (c) The computational workflow, including: (i) crystal structure prediction with the CHGGen generative model, (ii) thermodynamic stability evaluation using DFT and the Materials Project phase diagram, (iii) development of a fine-tuned MLIP for molecular dynamics simulations via active learning, and (iv) large-scale MD simulations with a fine-tuned CHGNet to derive Li diffusivities.  
        (d) Schematic of the CHGGen framework for generating \ce{LiPO2F2}. The process begins with unconditional generation, followed by the removal of Li atoms to create a refined, symmetrized polyanion framework. The framework is then used in an inpainting step to generate the final crystal structure.
        (e) Thermodynamic stability of the most stable \ce{LiPO2F2} polymorphs generated in each space group (blue bars: decomposition energy ($E_d$) predicted by pretrained CHGNet; orange bars: $E_d$ calculated from r$^2$SCAN-DFT). 
        }
    \label{fig:computation_framework}
\end{figure*}

In this work, we focus on the chemical composition \ce{LiPO2F2}, which has been established as a primary mixed anion electrolyte decomposition product in Li-ion batteries \cite{campion_thermal_2005, jayawardana_role_2021, parimalam_decomposition_2017, Spotte-Smith2023_Evan} and hence presents a possible SEI component. 

We combine diffusion-based generative models with machine-learning interatomic potential (MLIP) simulations to provide the first direct, atomistically resolved evidence that amorphous \ce{LiPO2F2} is a fast ion conductor with low interstitial Li$^+$ defect formation energies and rapid Li$^+$ diffusion. 
To enable a comparison between amorphous and crystalline states, we therefore predict the \ce{LiPO2F2} ground-state crystal structure and show that the amorphous phase exhibits low amorphorization energy, favorable Li$^+$ interstitial defect formation, and ionic conductivity orders of magnitude higher than the crystalline polymorph. 
These findings establish amorphous oxyfluorophosphates as possible single-ion conducting channels within the SEI, providing a fundamental explanation for the high performance of P- and F-containing batteries and offering a guiding principle for the molecular engineering of optimal battery interfaces.

\begin{figure*}[t]
    \centering
    \includegraphics[width=0.99\linewidth]{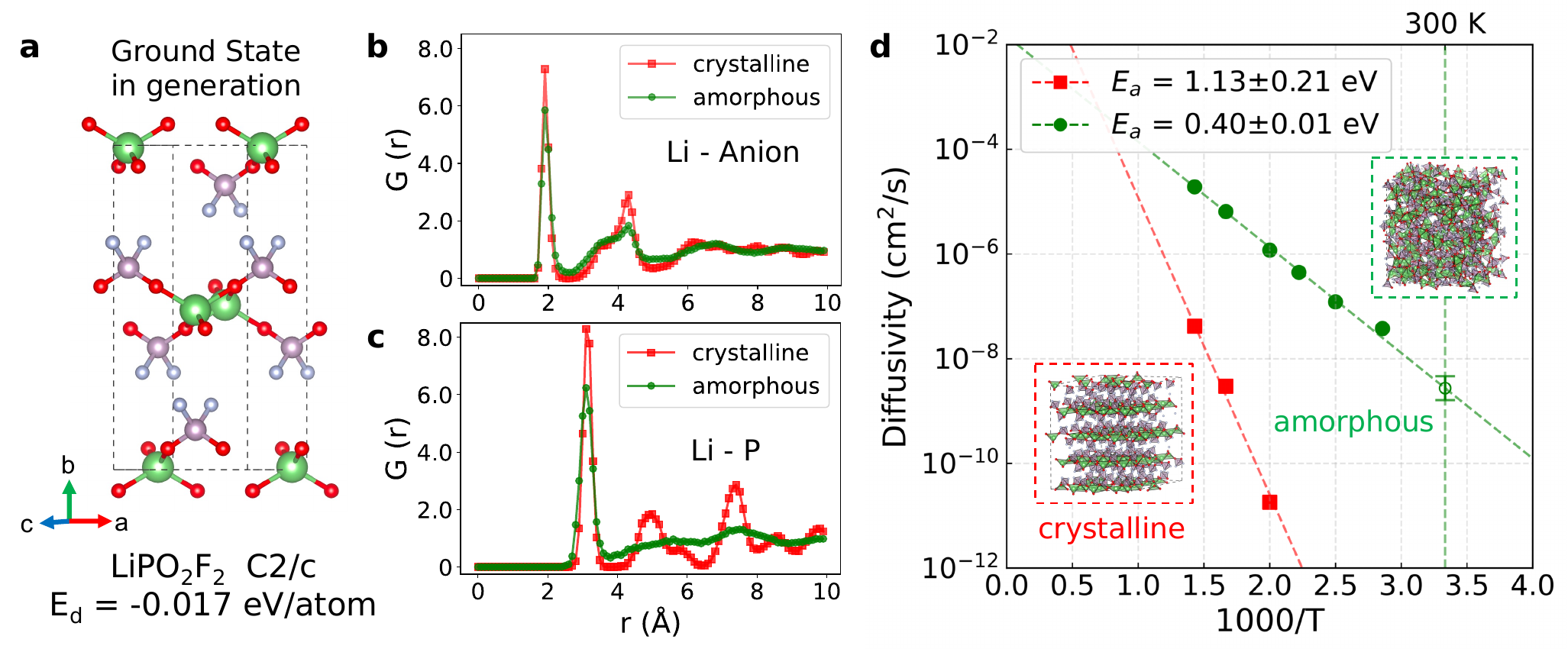}
    \caption{\textbf{Crystal structure and Li transport properties of \ce{LiPO2F2}.} (a) The predicted ground-state structure of \ce{LiPO2F2} ($C2/c$), which has a decomposition energy of $E_d = -0.017$ eV/atom relative to the r$^2$SCAN-DFT Materials Project phase diagram.
    (b, c) Radial distribution functions (RDFs) for the crystalline (orange) and amorphous (blue) phases at 300 K, showing (b) Li--anion (O/F) and (c) Li--P correlations.
    (d) Arrhenius plots of Li diffusivity from MD simulations for the crystalline (red squares) and amorphous (green dots) phases. Dashed lines are linear fits used to extract the activation energies ($E_a$).
    }
    \label{fig:diff_LiPOF}
\end{figure*}

\textit{Generation of structures:}
A crucial step in evaluating the role of \ce{LiPO2F2} as a primary SEI component is to identify a stable, low-energy crystalline structure in the condensed phase to serve as a reference for comparison with amorphous counterparts. Starting from the molecular representation in Figure~\ref{fig:computation_framework}a, the problem is conceptualized as a crystal structure prediction (CSP) task \cite{Zhou2025_drug, aykol_predicting_2024}. The computational molecular CSP is well-established, including structure searching, energy ranking with surrogate models, and final validation with quantum chemical calculations \cite{gharakhanyan_fastcsp_2025}. 
However, such methods rely on the packing of discrete molecules, where the specific orientation of the unit is critical and difficult to optimize for ionic systems. In the inorganic condensed phase, structure generation is better approached through processes that intrinsically resolve the cation-anion coordination environment, eliminating the need to explicitly specify polyanion orientations.

To address this challenge, we performed CSP using a deep generative model, Crystal-Host Guided Generation (CHGGen) \cite{zhong_crystal_2025}. CHGGen is a diffusion model, a probability-based ML technique that builds a crystal structure by starting with a random arrangement of atoms and progressively refining it into a chemically and physically stable configuration using score-based denoising \cite{song2020score, xie2021_cdvae, Zeni2025, Hsu2024}. 
The key feature of CHGGen lies in its unique two-stage strategy (Figure \ref{fig:computation_framework}d). 
The first unconditional generation process yields a prior structure with physically plausible features, such as local bonding environments. Given this prior, CHGGen temporarily removes the \ce{Li+} ions and refines the remaining \ce{PO2F2-} polyanion framework into a symmetrized structure with a higher space group.
Once the framework is established, the \ce{Li+} ions are reinserted into their most probable positions via a guided inpainting process \cite{lugmayr_repaint_2022}. 
This two-step process allows CHGGen to efficiently navigate the vast number of possible atomic arrangements and pinpoint highly-ordered \ce{LiPO2F2} crystal structures. For complete methodological details, we refer readers to Ref.~\cite{zhong_crystal_2025} and the Supporting Information (SI). 

After generative sampling, CHGGen employs the pretrained-CHGNet MLIP to rapidly explore the potential energy landscape of the generated structures \cite{Deng2023_chgnet}. All candidates were relaxed and screened for thermodynamic stability by computing their decomposition energies ($E_d$) relative to the Materials Project GGA/GGA+$U$ phase diagram \cite{Wang2021_compatability}. Low-energy structures ($E_d < 0.03$ eV/atom) were further refined with r$^2$SCAN DFT calculations, yielding updated $E_d$ values with respect to the r$^2$SCAN phase diagram in the Li–P–O–F chemical space \cite{Furness2020_r2SCAN, horton_accelerated_2025}. Figure \ref{fig:computation_framework}e presents the CHGNet (blue bar) and r$^2$SCAN-DFT (orange bar) calculated decomposition energies for the most stable polymorphs in each distinct space group. 
The pretrained-CHGNet potential identified six thermodynamically stable structures ($E_d\leq0$) and correctly predicted their relative energy ranking compared to DFT. The systematic underestimation of the absolute $E_d$ is a known effect of potential softening in universal MLIPs that favors smoother energy landscapes \cite{PES_soft}.

Following r$^2$SCAN-DFT calculations, we identified a candidate ground state as a C2/c polymorph with a DFT-calculated $E_d=-0.017$ eV/atom (Figure \ref{fig:diff_LiPOF}a). 
Structurally, this configuration is built from \ce{PO2F2} tetrahedra that are interconnected by corner-sharing \ce{LiO4} tetrahedra. This arrangement preserves the core \ce{PO2F2} motif observed in the molecular precursor, establishing a clear structural link between the gas-phase species and the condensed inorganic solid. The identified, condensed crystal structure also demonstrates a similar structural motif as the microporous structure of \ce{LiPO2F2} reported in Ref.~\cite{han_first_2019}.

Having identified a plausible candidate for the ground-state structure, the amorphous structures were generated via well-benchmarked melt-quench molecular dynamics (MD) \cite{aykol_thermodynamic_2018, sivonxay_density_2022} using a supercell $>10$ \AA\ in each direction. Modeling the structural disorder of amorphous materials requires simulations of large systems over long timescales, a task for which MLIPs are ideally suited due to their near-ab-initio accuracy at a fraction of the computational cost \cite{Qi2021_gap, Zhong2024_LZC}.
We fine-tuned the pretrained CHGNet model on a custom dataset of energies, forces, and stresses from DFT calculations, creating a specialized potential accurate for the Li--P--O--F chemical space (see SI) \cite{Vandermause2022_active, Xie2023_flare}. Using this fine-tuned CHGNet, the ground-state crystal was heated to 1000 K to induce a molten (or liquid) state, then quenched and equilibrated at target temperatures ranging from 300 K to 700 K. This approach ensures that the resulting amorphous structure is a reasonable, low-energy configuration rather than an arbitrary arrangement of atoms. The amorphous state was confirmed by comparing the radial distribution functions (RDFs) of the crystalline and amorphous phases (Figure \ref{fig:diff_LiPOF}b,c). While short-range Li--anion (O/F) coordination is mostly preserved, the loss of long-range Li--P correlation beyond 4 \AA\ is characteristic of an amorphous structure.

To evaluate the thermodynamic favorability of amorphization for \ce{LiPO2F2} versus other common SEI components, including LiF (mp-1138, the Materials Project ID), \ce{Li2O} (mp-1960), \ce{Li2CO3} (mp-3054), we calculated the amorphization energy $\Delta E_{\text{pot}}$. This term represents the energetic penalty of disordering the crystal, defined as $\Delta E_{\text{pot}} = \langle E^{\text{a}}_{\text{pot}} \rangle - \langle E^{\text{c}}_{\text{pot}} \rangle$, where $\langle E^{\text{a}}_{\text{pot}} \rangle$ and $\langle E^{\text{c}}_{\text{pot}} \rangle$ are the average potential energies of the amorphous and crystalline phases, respectively, sampled from equilibrium MD simulations at 300 K.
\begin{table}[tb]
    \centering
    \begin{tabularx}{\linewidth}{cccc}
        \toprule\midrule
        System &~ $\Delta E_{\text{pot}}$ [meV/atom] &~ $E_a$ [eV] &~ $\rho$ [g/cm$^3$] \\
        \midrule
        c-\ce{LiPO2F2} &~ -- &~ $1.13\pm0.21$ &~ $2.04\pm0.01$ \\
        \midrule
        a-\ce{LiPO2F2} &~ $30\pm2$ &~ $0.40\pm0.01$ &~ $1.96\pm0.01$ \\
        \midrule
        a-\ce{Li2CO3} &~ $64\pm2$ &~ $0.39\pm0.01$ &~ $1.97\pm0.01$ \\
        \midrule
        a-LiF &~ $95\pm 2$ &~ -- &~ $2.07\pm0.01$ \\
        \midrule
        a-\ce{Li2O} &~ $124\pm2$ &~ -- &~ $1.85\pm0.01$ \\
        \midrule\bottomrule
    \end{tabularx}
    \caption{
    The amorphization energy $\Delta E_{\text{pot}} = \langle E^{\text{a}}_{\text{pot}}\rangle - \langle E^{\text{c}}_{\text{pot}}\rangle $ compares the energy difference between amorphous (a-) and crystalline (c-) potential energies. $\Delta E_{\text{pot}}$ and density ($\rho$) are calculated from equilibrium MD simulations at 300 K. 
    $E_a$ is the Li diffusion activation energy. The $E_a$ for amorphous LiF and \ce{Li2O} is not reported due to their thermodynamic instability, which leads to crystallization in MD equilibrations at elevated temperatures. The standard deviation is derived based on the thermal fluctuations in MD simulations.
    }
    \label{tab:summary}
\end{table}
As shown in Table~\ref{tab:summary}, \ce{LiPO2F2} exhibits a relatively low amorphization energy ($\Delta E_{\text{pot}}$) of 30 meV/atom. This value is approximately half that of \ce{Li2CO3} (64 meV/atom) and significantly lower than that for LiF (95 meV/atom) and \ce{Li2O} (124 meV/atom). 
This low energy cost suggests that amorphization is more thermodynamically accessible for \ce{LiPO2F2} than for the other components. This finding is consistent with the experimentally observed SEI morphology: components with high amorphization penalties (LiF, \ce{Li2O}, \ce{Li2CO3}) form nanocrystallites embedded within the amorphous matrix \cite{li_atomic_2017, Han2021_cryoTEM}.

\textit{Intrinsic Li-ion diffusivity:}
Following equilibration, production MD simulations were performed to analyze the Li$^+$ diffusion dynamics. 
Li diffusivity at each temperature was calculated from the mean-squared displacement (MSD) of atomic trajectories, and the activation energy ($E_a$) for diffusion was extracted by fitting the results to the Arrhenius equation (Figure \ref{fig:diff_LiPOF}d) \cite{He_Zhu_Epstein_Mo_2018}. 
Because statistically significant Li-ion hopping events were rare at room temperature, the diffusivity value for the amorphous phase at 300 K (unfilled dot) was determined by extrapolation from the high-temperature data.

Crystalline \ce{LiPO2F2} exhibits a steep Arrhenius slope (red dashed line in Figure~\ref{fig:diff_LiPOF}d), corresponding to a high activation energy for Li diffusion ($E_a = 1.13 \pm 0.21$~eV). 
This substantial energy barrier effectively prevents interstitial Li migration at lower temperatures; indeed, no statistically significant hopping events were observed in our MD simulations until the structure exhibits pronounced thermal disorder above 800 K.
In contrast, amorphous \ce{LiPO2F2} demonstrates markedly different transport behavior, characterized by a significantly lower activation energy ($E_a=0.40 \pm 0.01$~eV). It is worth noting that a 60 meV change in activation energy corresponds to roughly an order of magnitude difference in Arrhenius diffusivity at room temperature. This results in a projected room-temperature ($T=300$~K) calculated ionic conductivity of $\sigma^{\text{calc}}_{\text{Li}} \approx 0.18$~mS cm$^{-1}$.

\begin{figure}[tb]
    \centering
    \includegraphics[width=0.95\linewidth]{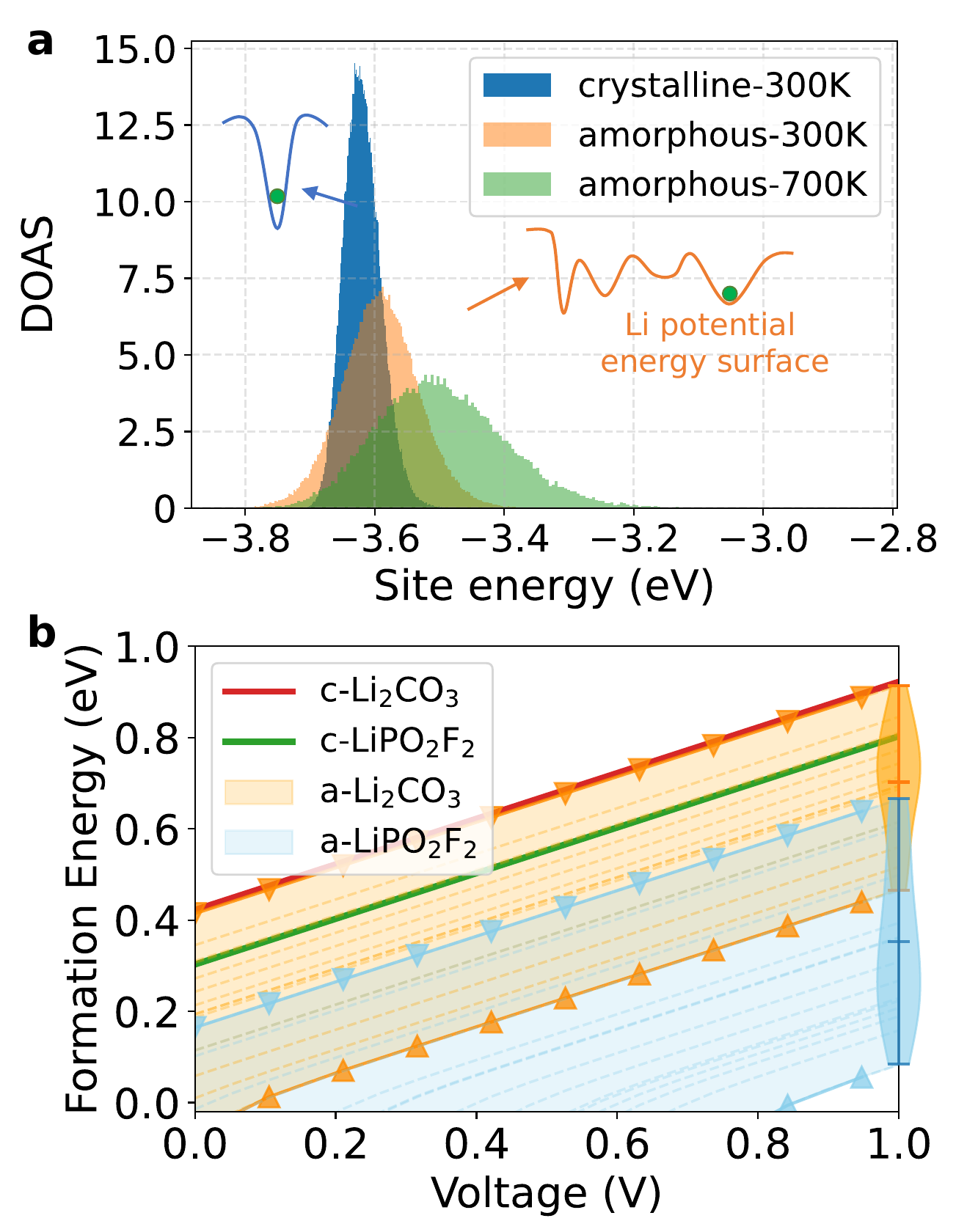}
    \caption{\textbf{Li site energy and interstitial defect formation energy distribution in \ce{LiPO2F2}.} 
    (a) Density of atomic states (DOAS) for Li ions in crystalline (blue) and amorphous (orange) \ce{LiPO2F2} at 300 K. The distribution for the amorphous phase at 700 K (green) illustrates thermal broadening of the energy landscape.
    (b) The calculated formation energy for a Li interstitial defect as a function of voltage (\textit{vs.}~Li/Li$^+$) in crystalline (c-) and amorphous (a-) phases. The solid red and green lines represent the c-\ce{Li2CO3} and c-\ce{LiPO2F2}, respectively. 
    The shaded regions, bounded by lines with triangles, represent the distribution of formation energies for an ensemble of defects in the amorphous phases; individual calculations are shown as dashed lines.
    The histograms on the right illustrate the distributions at 1 V.
    }
    \label{fig:DOAS}
\end{figure}

To understand the atomistic origin of the enhanced conductivity in amorphous \ce{LiPO2F2}, we analyzed the density of atomic states (DOAS) \cite{Wang2023_frustration}, which maps the distribution of site energies experienced by Li ions. 
As shown in Figure \ref{fig:DOAS}a, the Li site energy landscape in the crystalline phase (blue histogram) is characterized by a narrow distribution, indicating that Li ions occupy well-defined, deep potential wells. 
High ionic conductivity requires a broadened energy landscape where the energy cost for an ion to move between adjacent sites is minimal and can be overcome by thermal energy \cite{zeng_high-entropy_2022}. 
The DOAS analysis of amorphous \ce{LiPO2F2} (orange histogram) clearly shows that amorphization significantly broadens this landscape, creating a continuous network of energetically accessible sites.
This flattened energy profile rationalizes the dramatically lower activation energy and higher Li diffusivity observed in the amorphous phase.

\textit{Li-ion defect formation energy:}
Beyond intrinsic diffusion, extrinsic mechanisms (e.g., the formation of charge-carrying defects) can enhance ionic conductivity in the SEI. Near the Li-metal anode, this is dominated by the creation of Li interstitials through a ``Li-stuffing" process, where a low defect formation energy enables the host material to accommodate excess Li \cite{zhang_kinetic_2020, xiao_lithium_2021, chen_unlocking_2024}. The distinction between defect formation and ion migration is critical.  For instance, \citet{shi_defect_2013} revealed that the migration barrier for an interstitial Li is only 0.24~eV in crystalline \ce{Li2CO3}, but its formation energy is a prohibitive 0.3--0.8 eV in the low-voltage range. In crystalline LiF,  the most common defects, vacancies of F and Li, exhibit even higher formation energies in the range of 0.76 to 1.3~eV \cite{yildirim_first-principles_2015}. As a result of such high formation energies, the equilibrium defect concentrations remain very low $\approx 10^{-6}$ to $10^{-10}$ cm$^{-3}$ at 0~V \textit{vs.} Li/Li$^+$, which is 4--8 orders of magnitude lower than the levels typically required for efficient defect-mediated ion transport. To clarify the charge-carrier concentration limitations in key SEI components, we performed defect formation energy calculations for both crystalline and amorphous \ce{Li2CO3}, as well as for our amorphous and predicted crystalline \ce{LiPO2F2}. Candidate interstitial Li-ion sites are identified based on the static calculation of the charge density distribution of the undoped structure \cite{shen_charge-density-based_2020}. For amorphous structures, sites were randomly selected to sample different coordination environments. The interstitial defect formation energy was calculated at charge state $q=1$ with the Freysoldt correction (see SI) \cite{freysoldt_fully_2009, farnell_ab_2025}.

Figure \ref{fig:DOAS}b compares the Li$^+$ interstitial defect formation energy as a function of voltage. 
The solid red and green lines represent the crystalline \ce{Li2CO3} and \ce{LiPO2F2} phases, where the c-\ce{LiPO2F2} shows $\sim$0.1 eV decrease over c-\ce{Li2CO3}.
The shaded orange and blue regions in Figure~\ref{fig:DOAS}b display the distribution of formation energies for the corresponding amorphous phases. 
Upon structural amorphorization, as expected, the defect formation energy exhibits significant variability \cite{farnell_ab_2025}, where the mean value drops for both \ce{Li2CO3} and \ce{LiPO2F2}. 
Here, we only compare the trends in the positive range. The mean value of defect formation energy at 1 V decreases from $0.7$ eV (a-\ce{Li2CO3}) to $0.35$ eV (a-\ce{LiPO2F2}).
This low formation energy, combined with the fact that Li-P-O-F phases have been observed as amorphous within the SEI \cite{nguyen2025_insitu}, suggests that the amorphous \ce{LiPO2F2} is well positioned to enable fast Li transport in the SEI.

\textbf{Discussion} -- 
A persistent challenge in designing SEI functionality lies in identifying the enabling components within the mosaic surface layer, first suggested by \citet{peled_reviewsei_2017}. 
For example, the robustly insulating nature of LiF provides excellent electronic passivation, yet is unlikely to enable high-performance single-ion conductivity. 
Although defect-mediated mechanisms can improve Li diffusion \cite{de_angelis_exploring_2025}, the high defect energy (0.76 eV for Li$^+$ vacancies and 1.73 eV for Li$^+$ interstitials  \cite{yildirim_first-principles_2015, de_angelis_exploring_2025}) precludes a major contribution from bulk LiF, as evidenced by the inferior rate capability of SEIs grown entirely from LiF \cite{he_intrinsic_2020}. Similarly, recent work on fabricated single-component `SEI'-like \ce{Li2O} films shows that such films exhibit an ionic conductivity of ${\sim}10^{-9}$ S cm$^{-1}$ with an activation energy of ${\sim}0.47$ eV \cite{guo_li_2020}.

While decades of work have attempted to explain the inner SEI conductivity through its proposed nanocrystalline domains (\ce{Li2O}, \ce{Li2CO3}, and \ce{LiF}), other efforts have focused on transport occurring at grain boundaries between these crystalline domains. 
Atomistic simulations have revealed, for instance, that the LiF/\ce{Li2O} interface can harbor a high concentration of charge carriers with enhanced ionic conductivity \cite{ma_origin_2022}. 
Similarly, studies on the LiF/\ce{Li2CO3} interface have shown an accumulation of ionic carriers, leading to increased ionic transport \cite{pan_design_2016}. 
More recent work employing MLIPs has suggested that high Li-ion conductivity can originate from an amorphous mixture of LiF and \ce{Li2CO3} \cite{Hu2023_jacs}. 
These studies hint at the importance of grain boundaries and, by extension, possibly larger inorganic amorphous domains.

While we do not rule out the possibility of other functionality-enhancing SEI components (e.g., LiF-LiH solid solution \cite{liu_probing_2025}), our work suggests a complementary and perhaps dominant mechanism: that amorphous, mixed-anion LiPO$_x$F$_y$ phases, formed from electrolyte decomposition, function as the primary Li-conducting channels throughout the SEI matrix. The activation barrier of amorphous \ce{LiPO2F2} ($E_a = 0.40$ eV) exceeds that of typical superionic conductors (0.1–0.3 eV) \cite{Jun2025_Matter}, however, it remains sufficient for ion transport across an SEI layer of nanoscale thickness  (10–50 nm)~\cite{li_atomic_2017}. Importantly, amorphous \ce{LiPO2F2} shows excellent energetics for Li$^{+}$ defect formation, especially in comparison to crystalline phases such as LiF, \ce{Li2CO3}, \ce{Li2O}, and even amorphous \ce{Li2CO3}.  Although our study highlights a specific lithium difluorophosphate composition, conductivity is anticipated to evolve smoothly by stoichiometric tuning within similar coordination environments, increasing as the fluorine content rises (see SI) \cite{cheng_evaluation_2020}. It suggests that a broader class of amorphous lithium fluorophosphates can serve as ion-conducting domains and interparticle ``freeways" that facilitate transport through the inorganic SEI. Supporting our hypothesis, \citet{Zhang2025_NM} showed that additives such as trifluoroacetic anhydride (TFAA) and diphosphoryl fluoride (DPF) undergo preferential sacrificial reduction during initial cycling, generating LiPO$_x$F$_y$ species that yield a more stable and conductive interphases. 
Similarly, \citet{ren_densification_2024} reported that Zr–O surface deposits promote the densification of the cathode–electrolyte interphase (CEI) on \ce{LiCoO2}, where enrichment of \ce{LiPO2F2} and its transformation into mixed-anion phases (ZrO$_x$F$_y$ and LiZrO$_x$F$_y$) significantly enhanced Li-ion transport kinetics. 
Taken together, the evidence calls for increased attention to such mixed-anionic species, in particular, in disordered and amorphous regions of the SEI, and their distribution and connectivity throughout the SEI. 

In summary, we combine diffusion-based generative models with MLIP-driven simulations to investigate the structure and Li transport properties of \ce{LiPO2F2}, a model SEI component. We show that the amorphous structure features favorable defect energetics and rapid ionic transport, identifying it as a strong candidate for Li-ion conduction within the SEI of Li-ion batteries.
These findings offer an atomistic rationale for the superior performance of fluorophosphate-rich SEIs, highlighting amorphous LiPO$_x$F$_y$ phases as key ion-conducting pathways and providing a mechanistic foundation for improving battery performance via molecular design and interfacial engineering strategies \cite{Yu2022_molecule_design, Holoubek2020, wang_data-driven_2025}.

\textbf{Acknowledgements} -- 
This work was intellectually led by the Battery Materials Research program under the Assistant Secretary for Energy Efficiency and Renewable Energy, Office of Vehicle Technologies of the U.S. Department of Energy, Contract DE-AC0205CH11231. The computations were supported by the National Energy Research Scientific Computing Center (NERSC) under the GenAI Project and the National Renewable Energy Laboratory (NREL) clusters under the silimorphous allocation. P.Z. acknowledges funding support from the BIDMaP Postdoctoral Fellowship. The authors thank Solomon Oyakhire, Weiran Zhang, and Bingqing Cheng for valuable discussions.

\bibliography{references}
\end{document}